\documentclass[aps,preprint,amsmath,amssymb]{revtex4}

\usepackage{graphicx}
\usepackage{dcolumn}
\usepackage{bm}

\def\beq{\begin{eqnarray}}
\def\eeq{\end{eqnarray}}

\def\GeV{{\rm GeV} }

\def\non{\nonumber}

\def\la{\langle}
\def\ra{\rangle}

\def\eff{\rm eff}

\newcommand\npb{Nucl.\ Phys.\ B }
\newcommand\npps{Nucl.\ Phys.\ B (Proc.\ Suppl.) }
\newcommand\plb{Phys.\ Lett.\ B }
\newcommand\zpc{Z.\ Phys.\ C }

\begin{document}
\title{\Large \bf Factorization and polarization in two charmed-meson B decays}
\author{\large \bf
Chuan-Hung~Chen$^{a,d}$ \footnote{Email: phychen@mail.ncku.edu.tw},
Chao-Qiang~Geng$^{b}$ \footnote{Email: geng@phys.nthu.edu.tw} and
Zheng-Tao~Wei$^{c,d}$ \footnote{Email: weizt@phys.sinica.edu.tw } }
\affiliation{
$^{a}$Department of Physics, National Cheng-Kung University, Tainan, 701 Taiwan \\
$^{b}$Department of Physics, National Tsing-Hua University, Hsin-Chu, 300 Taiwan \\
$^{c}$ Institute of Physics, Academia Sinica, Taipei, 115 Taiwan \\
$^{d}$ National Center for Theoretical Sciences, Taiwan }

\begin{abstract}

We provide a comprehensive test of factorization in the heavy-heavy
$B$ decays motivated by the recent experimental data from BELLE and
BABAR collaborations. The penguin effects are not negligible in the
$B$ decays with two pseudoscalar mesons. The direct CP asymmetries
are found to be a few percent. We give estimates on the weak
annihilation contributions by analogy to the observed
annihilation-dominated processes. The $N_c$ insensitivity of
branching ratios indicates that the soft final state interactions
are not dominant. We also study the polarizations in $B\to
D^*D_{(s)}^*$ decays. The power law shows that the transverse
perpendicular polarization fraction is small. The effects of the
heavy quark symmetry breaking caused by the perturbative QCD and
power corrections on the transverse polarization are also
investigated.

\end{abstract}
\maketitle

\section{Introduction}

The study of $B$ meson weak decays is of high interest in heavy
flavor physics and CP violation. In particular, much attention has
been paid to the two-body charmless hadronic $B$ decays, but there
are relatively less discussions on the decays with charmful mesons,
such as the modes with two charmed-meson in which the final states
are both heavy. However, the two charmed-meson decays can provide
some valuable and unique information which is different from the
light meson productions. For example, CP asymmetries in the decays
of $B\to D^{(*)+}D^{(*)-}$ play important roles in testing the
consistency of the standard model (SM) as well as exploring new
physics \cite{SX}. Moreover, these decays are ideal modes to check
the factorization hypothesis as the phenomenon of color transparency
for the light energetic hadron is not applicable. Since the decay
branching ratios (BRs), CP asymmetries (CPAs) and polarizations of
$B\to D^{(*)} D^{(*)} (D^{(*)} D^{(*)}_s )$ have been partially
observed in experiments \cite{PDG04, BELLE_DD, BABAR_DD}, it is
timely to examine these heavy-heavy $B$ decays in more detail.

At the quark level, one concludes that the two charmed-meson decays
are dominated by tree contributions since the corresponding
inclusive modes are $b\to q c\bar{c}$ with $q=s$ and $d$. It is
known that the factorization has been tested to be successful in the
usual color-allowed processes. However, the mechanism of
factorization in heavy-heavy decays is not the same as the case of
the light hadron productions. The color transparency argument
\cite{Bjorken} for light energetic hadrons is no longer valid to the
modes with heavy-heavy final states. The reason can be given as
follows. Due to the intrinsic soft dynamics in the charmed-meson,
non-vanishing soft gluon contributions are involved in the strong
interactions between an emission heavy meson with the remained
$BD^{(*)}$ system. Since the corresponding divergences may not be
absorbed in the definition of the hadronic form factor or hadron
wave function, the decoupling of soft divergences is broken. This
means that the mechanism of factorization has to be beyond the
perturbative frameworks, such as QCD factorization \cite{BBNS} and
soft-collinear effective theory \cite{SCET}. The large $N_c$ limit
is another mechanism to justify factorization \cite{BGR},
corresponding to the effective color number $N_c=\infty$ in the
naive factorization approach \cite{BSW}. The understanding of
factorization in heavy-heavy decays requires some quantitative
knowledge of non-perturbative physics which is not under control in
theory. In this paper, we will assume the factorization hypothesis
and apply the generalized factorization approach (GFA) \cite{GFA1,
GFA2} to calculate the hadronic matrix elements.

It is known that annihilation contributions and nonfactorizable
effects with final state interactions (FSIs) play important role
during the light meson production in $B$ meson decays. For instance,
to get large strong phases for CP asymmetries (CPAs) in $B\to K \pi$
and $B\to \pi \pi$ decays, these effects are included inevitably
\cite{KLS_PRD63, CCS_PRD71}. Moreover, they are also crucial to
explain the anomaly of the polarizations in $B\to \phi K^*$ decays,
measured by BABAR \cite{babar_pol} and BELLE \cite{belle_pol}
recently. By the naive analysis in flavor diagrams, one can easily
see that the decay modes of $B\to D^{(*)0} \bar{D}^{(*)0}$ and
$D^{(*)0}_{s} \bar{D}^{(*)0}_{s}$ are annihilation-dominated
processes. Therefore, measurements of these decays will clearly tell
us the importance of annihilation contributions in the production of
two charmed-meson modes.

For the color-allowed decays, since the short-distant (SD)
nonfactorizable parts are associated with the Wilson coefficient
(WC) of $C_1/N_c$, where $C_1$ is induced by the gluon-loop and it is
much smaller than $C_2\sim 1$, we can see that the effects arising
from the SD non-spectator contributions should be small \cite{LM}.
Nevertheless, long-distant (LD) nonfactorizable contributions
governed by rescattering effects or FSIs may not be negligible.
Inspired by the anomaly of the large transverse perpendicular
polarization, denoted by $R_{\perp}$, in $B\to \phi K^*$ decays, if
there exist significant LD effects, we believe that large values of
$R_{\perp}$ may appear in $B\to D^{*} D^{*}$ and $B\to D^{*}
D^{*}_s$ too.
As we will discuss, the power-law in the two-vector
charmed-meson decay leads to a small $R_{\perp}$. The recent
measurement of the polarization fraction by the BELLE collaboration
gives $R_{\perp}=0.19\pm 0.08\pm 0.01$ \cite{BELLE_DD} in which the central
value is about a factor of three comparing with the model-independent
prediction within the factorization approach and heavy quark symmetry.
Clearly, to get the implication from the data, we need a detailed
analysis in two charmful final states of $B$ decays.

To estimate the relevant hadronic effects for two-body decays in the
$B$ system, we use the GFA, in which the leading effects are
factorized parts, while the nonfactorized effects are lumped and
characterized by the effective number of colors, denoted by
$N^{\eff}_c$. Note that the scale and scheme dependences in
effective WCs $C^{\eff}_{i}$ are insensitive. In the framework of
the GFA, since the formulas for decay amplitudes are associated with
the transition form factors, we consider them based on heavy quark
effective theory (HQET) \cite{NeubertHQET}. We will also study their
$\alpha_s$ \cite{NeubertQCD} and power corrections which break heavy
quark symmetry (HQS) \cite{NR}. In our analysis, we will try to find
out the relationship between the HQS and its breaking effects for
$R_{\perp}$. In addition, we will reexamine the influence of penguin
effects, neglected in the literature \cite{LR}. We
will show that sizable CPAs in $\bar B^0 \to D^+ D^-$ and $B^-\to
D^{0} D^{-}$ decays may rely on FSIs.

This paper is organized as follows. In Sec. II, we give the
effective Hamiltonian for the heavy-heavy B decays. The definition
of heavy-to-heavy form factors are also introduced. In Sec. III, we
show the general formulas for B to two charmful states in
the framework of the GFA. The effects of the heavy quark symmetry
breaking on the transverse perpendicular polarization are
investigated. In Sec. IV, we provide the numerical predictions on
the BRs, direct CPAs and the polarization fractions. Conclusions are
given in Sec. V. We collect all factorized amplitudes for $B\to PP,~
PV(VP)$ and $VV$ decays in Appendixes.

\section{ Effective interactions and parametrization of form factors}

The relevant effective Hamiltonian for the $B$ meson decaying to two
charmful meson states is given by \cite{BBL},
\begin{eqnarray}
H_{\rm eff}(\Delta B=1)&=&\frac{G_F}{\sqrt 2}\left\{
V_{ub}V_{uq}^{*}[C_1(\mu)O^{u}_1+ C_2(\mu)O^{u}_2]
+V_{cb}V_{cq}^{*}[C_1(\mu)O^{c}_1+ C_2(\mu)O^{c}_2] \right.\nonumber
\\
&& \left. -V_{tb}V_{tq}^{*}\sum_{k=3}^{10}C_k(\mu)O_k\right\}+H.c.
\;, \label{effj}
\end{eqnarray}
where $q=s$ and $d$, $V_{ij}$ denote the Cabibbo-Kobayashi-Masikawa
(CKM) matrix elements, $C_{i}(\mu)$ are the Wilson coefficients
(WCs) and  $O_i$ are the four-fermion operators, given by
\begin{eqnarray}
& &O^{u}_1 = (\bar{q}_iu_j)_{V-A}(\bar{u}_jb_i)_{V-A},
\;\;\;\;\;\;\;\; O^{u}_2 = (\bar{q}_iu_i)_{V-A}(\bar{u}_jb_j)_{V-A},
\nonumber \\
& &O^{c}_1 = (\bar{q}_ic_j)_{V-A}(\bar{c}_jb_i)_{V-A},
\;\;\;\;\;\;\;\;\; O^{c}_2 =
(\bar{q}_ic_i)_{V-A}(\bar{c}_jb_j)_{V-A},\nonumber
\end{eqnarray}
\begin{eqnarray}
& &O_{3(5)}
=(\bar{q}_ib_i)_{V-A}\sum_{q^{\prime}}(\bar{q}^{\prime}_jq^{\prime}_j)_{V\mp
A}, \;\;\;\;\;\;\;\;
O_{4(6)}=(\bar{q}_ib_j)_{V-A}\sum_{q^{\prime}}(\bar{q}^{\prime}_jq^{\prime}_i)_{V\mp
A},
\nonumber \\
& &O_{7(9)}
=\frac{3}{2}(\bar{q}_ib_i)_{V-A}\sum_{q^{\prime}}e_{q^{\prime}}
 (\bar{q}^{\prime}_jq^{\prime}_j)_{V\pm A}, \;\; O_{8(10)}
=\frac{3}{2}(\bar{q}_ib_j)_{V-A}\sum_{q^{\prime}}e_{q^{\prime}}
 (\bar{q}^{\prime}_jq^{\prime}_i)_{V\pm A}.
\end{eqnarray}
with $i$ and $j$ being the color indices, $O_{3-6}$ ($O_{7-10}$)
 the QCD
(electroweak) penguin operators
and $(\bar{q}_{1} q_{2})_{V\pm A}=\bar{q}_{1} \gamma_{\mu} (1\pm
\gamma_5 ) q_2$.
In order to cancel the renormalization scale and
scheme dependence in the WCs of $C_{i}(\mu)$, the effective WCs are
introduced by %
\beq %
C(\mu)\la O(\mu) \ra \equiv C^{\eff}\la O \ra_{\rm tree}.
\eeq %
Since the matrix element $\la O \ra_{\rm tree}$ is given at tree
level, the effective WCs are renormalization scale and scheme
independent. To be more useful, we can define the effective WCs as
\begin{eqnarray}
a^{\rm eff}_{1} &=& C_2^{\eff} +
  \frac{C_{1}^{\eff}}{(N_c^{\eff})_1}, ~~~~~~
a^{\rm eff}_2 = C_1^{\eff} +
  \frac{C_2^{\eff}}{(N_c^{\eff})_2},  \non \\
a^{\rm eff(q)}_{3,4} &=& C_{3,4}^{\eff} +
  \frac{C_{4,3}^{\eff}}{(N_c^{\eff})_{4,3}}+
  \frac{3}{2}e_{q}\left( C_{9,10}^{\eff}+
  \frac{C_{10,9}^{\eff}}{(N_c^{\eff})_{10,9}} \right), \nonumber \\
a^{\rm eff(q)}_{5,6} &=& C_{5,6}^{\eff}+
  \frac{C_{6,5}^{\eff}}{(N_c^{\eff})_{6,5}}+
  \frac{3}{2}e_q \left(C_{7,8} +
  \frac{C_{8,7}^{\eff}}{(N_c^{\eff})_{8,7}} \right),
\label{effwc}
\end{eqnarray}
where %
\beq %
\frac{1}{(N_c^{\eff})_i}\equiv \frac{1}{N_c}+\chi_i. %
\eeq %
with $\chi_i$ being the non-factorizable terms. In the GFA,
$1/(N_c^{\eff})_i$ are assumed to be universal and real in
the absence of FSIs. In the naive factorization, all effective
WCs $C_i^{\eff}$ are reduced to the corresponding WCs of $C_i$ in the
effective Hamiltonian and the non-factoziable terms are neglected,
i.e., $\chi_i=0$.

Under the factorization hypothesis, the tree level hadronic matrix
element $\la O \ra_{\rm tree}$ is factorized into a product of two
matrix elements of single currents, which are represented by the decay
constant and form factors. The $B\rightarrow H\left( H=D,\
D^{*}\right)$ transition form factors are crucial ingredients in the
 GFA for the heavy-heavy decays. To obtain the transition elements of
$B\rightarrow H$ with various weak vertices, we first parameterize
them in terms of the relevant form factors under the conventional
forms as follows:
\begin{eqnarray}
\langle D| V_{\mu} | \bar{B} \rangle &=&
F_{1}(q^2)\Big\{P_{\mu}-\frac{P\cdot q }{q^2}q_{\mu} \Big\}
+\frac{P\cdot q}{q^2}F_{0}(q^2)\,q_{\mu}, \label{ffp} \\
\langle D^{*}(\epsilon )| V_{\mu} | \bar{B}%
\rangle &=&\frac{V(q^{2})}{m_{B}+m_{D^{*}}}\varepsilon _{\mu
\alpha \beta \rho }\epsilon ^{*\alpha }P^{\beta }q^{\rho },  \nonumber \\
\langle D^{*}(\epsilon )| A_{\mu} | \bar{B}
\rangle &=&i\left[2m_{D^{*}}A_{0}(q^{2})\frac{\epsilon ^{*}\cdot q}{%
q^{2}}q_{\mu }+( m_{B}+m_{D^{*}}) A_{1}(q^{2})\Big( \epsilon
_{\mu }^{*}-\frac{\epsilon ^{*}\cdot q}{q^{2}}q_{\mu }\Big)  \right. \nonumber \\
&&\left.-A_{2}(q^{2})\frac{\epsilon ^{*}\cdot q}{m_{B}+m_{D^{*}}}\Big( P_{\mu }-%
\frac{P\cdot q}{q^{2}}q_{\mu }\Big)\right],  \label{ffv}
\end{eqnarray}
where $V_{\mu}=\bar{q}\gamma _{\mu } b$, $A_{\mu}=\bar{q}\gamma
_{\mu } \gamma_{5} b$, $m_{B,D,D^*}$ are the meson masses,
$\epsilon_{\mu}$ denotes the
polarization vector of the $D^*$ meson, $P=p_{B}+p_{D^{(*)}}$,
$q=p_{B}-p_{D^{(*)}}$ and $P \cdot q=m^{2}_{B}-m^{2}_{D^{(*)}}$.
 According to the HQET, it will be more convenient to
define the form factors
in terms of
the velocity of the heavy quark rather than the momentum. The definition
of these form factors can be found in Ref. \cite{NR} and the relation
with the conventional ones are given by
\beq %
F_1(q^2)&=&\frac{m_B+m_D}{2\sqrt{m_Bm_D}}\left [\xi_+(w)
 -\frac{m_B-m_D}{m_B+m_D}~\xi_-(w) \right ], \non\\
F_0(q^2)&=&
 \frac{m_B+m_D}{2\sqrt{m_Bm_D}} \zeta_{D}(q^2)\left [\xi_+(w)
 -\frac{m_B+m_D}{m_B-m_D}\left (\frac{w-1}{w+1}\right )
 \xi_-(w) \right ], \non\\
V(q^2)&=&\frac{m_B+m_{D^*}}{2\sqrt{m_Bm_{D^*}}}~\xi_V(w), \ \ \
A_1(q^2)=
 \frac{m_B+m_{D^*}}{2\sqrt{m_Bm_{D^*}}}\zeta_{D^*}~\xi_{A_1}(w), \non\\
A_2(q^2)&=&\frac{m_B+m_{D^*}}{2\sqrt{m_Bm_{D^*}}}\left[\xi_{A_1}(w)
 +\frac{m_{D^*}}{m_B}~\xi_{A_2}(w) \right ], \non\\
A_3(q^2)&=&\frac{m_B+m_{D^*}}{2\sqrt{m_Bm_{D^*}}} \Bigg\{
 \frac{m_B}{m_B+m_{D^*}}(w+1)\xi_{A_1}(w)  \non \\
&& ~~~~~~~~~~~~~~~-\frac{m_B-m_{D^*}}{2 m_{D^*}}\left[
 \xi_{A_3}(w)+\frac{m_{D^*}}{m_B}~\xi_{A_2}(w)
 \right ]\Bigg \}, \non\\
A_0(q^2)&=&A_3(q^2)+\frac{q^2}{4m_Bm_{D^*}}\sqrt{\frac{m_B}{m_{D^*}}}
 \left [\xi_{A_3}(w)-\frac{m_{D^*}}{m_B}~\xi_{A_2}(w)\right ].
 \label{HQFF}
\eeq %
with $\omega=(m^{2}_B+m^{2}_{H}-q^2)/(2m_{B}m_{H})$ and
$\zeta_{H}(q^2)=1-q^2/(m_B+m_H)^2$. It is known that under the HQS,
$\xi_+=\xi_V=\xi_{A_1}=\xi_{A_3}=\xi(w)$ and $\xi_-=\xi_{A_2}=0$. In
our numerical estimations, we will base on the results of the HQS and
include $\alpha_s$ and $1/m_{B}$ power corrections as well.

\section{ Generalized Factorization formulas and polarization
 fractions of VV modes}

By the effective interactions and the form factors defined in the
previous chapter, the decay amplitude could be described by the
product of the effective WCs and the hadronic matrix elements in the
framework of the GFA. For the hadronic matrix elements
in $B\to PP$ decays, we will follow
the notation of Ref. \cite{GFA2}, given by
\begin{eqnarray}
X^{(BP_1,P_2)}_{1}&\equiv& \la P_2|(\bar{q}_{2} q_3)_{V-A}|0\ra
 \la P_1|(\bar{q}_{1} b)_{V-A}|\bar{B}\ra=if_{P_2}
 (m^{2}_{B}-m^{2}_{P_1})F^{BP_{1}}_{0}(m^{2}_{P_2}),  \non \\
X^{(BP_1,P_2)}_{2}&\equiv& \langle P_2|(\bar{q}_{2}
 q_3)_{S+P}|0\rangle \langle P_1|(\bar{q}_{1} b)_{S-P}|\bar{B}\rangle
 =-i\frac{m^{2}_{P_2}}{m_{2}+m_{3}}f_{P_2}
 \frac{m^{2}_{B}-m^{2}_{P_1}}{m_b-m_1}F^{BP_{1}}_{0}(m^{2}_{P_2}),~~~~
 \label{xpp}
\end{eqnarray}
 where
$(\bar{q}_1b)_{S-P}=\bar{q}_{1}(1-\gamma_5) b$, $(\bar{q}_2
q_3)_{S+P}=\bar{q}_{2}(1+\gamma_5) q_3$ and $m_{b,1,2,3}$ correspond to
the masses of quarks. The vertex $(S-P)\otimes (S+P)$ is from the
Fierz transformation of $(V-A)\otimes (V+A)$. To get the decay
constant and form factors for scalar vertices, we have utilized
equation of motion for on-shell quarks.
Moreover, we use
\begin{eqnarray}
X^{(BP,V)}_{1} &\equiv & \langle V| (\bar{q}_2 q_3)_{V-A}|0\rangle
\langle P|(\bar{q}_1 b)_{V-A}|\bar{B} \rangle=2f_V\,m_V F_1^{ B
P}(m_{V}^2)(\varepsilon^*\cdot p_{_{B}}),   \non \\
 X^{( BV,P)}_{1} &\equiv
& \langle P | (\bar{q}_2 q_3)_{V-A}|0\rangle \langle
V|(\bar{q}_1b)_{V-A}|\bar{B}\rangle=2f_P\,m_V A_0^{ B
V}(m_{P}^2)(\varepsilon^*\cdot p_{_{B}}), \non \\
X^{( BV,P)}_{2} &\equiv & \langle P | (\bar{q}_2 q_3)_{S+P}|0\rangle
\langle V|(\bar{q}_1b)_{S-P}|\bar{B}\rangle=\frac{2
m^{2}_{P}}{m_2+m_3}f_P\frac{m_V}{m_b+m_1} A_0^{ B
V}(m_{P}^2)(\varepsilon^*\cdot p_{_{B}}),~~~~\label{xpv}
\end{eqnarray}
 and
\begin{eqnarray}
 X^{( BV_1,V_2)} &\equiv & \langle V_2 | (\bar{q}_2
 q_3)_{V-A}|0\rangle \langle
V_1|(\bar{q}_1b)_{V-A}|\bar{B} \rangle \nonumber \\
&=& - if_{V_2}m_{V_2}\Bigg[ (\varepsilon^*_1\cdot\varepsilon^*_2)
(m_{B}+m_{V_1})A_1^{ BV_1}(m_{V_2}^2) - (\epsilon^*_1\cdot
p_{_{2}})(\epsilon^*_2 \cdot p_{_{1}}){2A_2^{ BV_1}(m_{V_2}^2)\over
(m_{B}+m_{V_1}) }  \nonumber \\ &&+
i\epsilon_{\mu\nu\alpha\beta}\varepsilon^{*\mu}_2\varepsilon^{*\nu}_1p^\alpha_{_{2}}
p^\beta_1\,{2V^{ BV_1}(m_{V_2}^2)\over (m_{B}+m_{V_1}) }\Bigg],
\label{hvv}
\end{eqnarray}
for $B\to PV(VP)$ and
 $B\to VV$, respectively.
We note that the sign difference of $X^{(BP_{1},P_2)}_{1}$ and
$X^{(BP_{1},P_2)}_{2}$ in Eq. (\ref{xpp}) will make the penguin
effects become non-negligible.
On the other hand,
the same sign of
$X^{(BV,P)}_{1}$ and $X^{(BV,P)}_{2}$ in Eq. (\ref{xpv}) leads to
the penguin effects negligible. Since the time-like form factors for
annihilation contributions are uncertain, we take $
Y^{(B,M_1M_2)}_{1(2)}\equiv \langle M_1 M_2|(\bar{q}_2 q_3)_{V\mp
A}|0\rangle \, \langle 0 | (\bar{q}_1 b)_{V-A}|\bar{B}\rangle$ and
$Y^{(B,M_1M_2)}_{3} \equiv \langle M_1 M_2|(\bar{q}_2
q_3)_{S+P}|0\rangle \, \langle 0 | (\bar{q}_1
b)_{S-P}|\bar{B}\rangle$ to represent them, where $M$ can be
pseudoscalar or vector bosons. Note that due to the identity of
$\varepsilon_{i}(p_{i})\cdot p_{i}=0$, we have used $\langle
V|(\bar{q}_{2} q_3)_{S+P}|0 \rangle =0$. With these notations
and associated effective WCs, one can display the
decay amplitude for the specific decay mode. We summary the relevant
decay amplitudes in Appendixes. Once we get the decay amplitude,
denoted by $A(B\to M_1 M_2)$, the corresponding decay rate of the
two-body mode could be obtained by
 \begin{eqnarray}
   \Gamma(B\to M_1 M_2)=\frac{G_F p}{16\pi m^{2}_{B}}|A(B\to M_1
   M_2)|^2.
 \end{eqnarray}
with $p$ being the spatial momentum of $M_{1,2}$.  Consequently, the
direct CPA is defined by
\begin{eqnarray}
A_{CP}=\frac{\bar{\Gamma}(\bar{B}\to \bar{M}_1
\bar{M}_2)-\Gamma(B\to M_1 M_2)}{\bar{\Gamma}(\bar{B}\to \bar{M}_1
\bar{M}_2)+\Gamma(B\to M_1 M_2)}. \label{cpa}
\end{eqnarray}

Besides the BRs and CPAs, we can also study the
polarizations of the vector mesons in $B\to VV$ decays. To discuss the
polarizations,
one can write the general decay
amplitudes
 in the helicity basis to be
\begin{eqnarray}
A^{(\lambda)}
&=&\epsilon_{1\mu}^{*}(\lambda)\epsilon_{2\nu}^{*}(\lambda) \left[ a
\, g^{\mu\nu} +\frac{b}{m_{V_1} m_{V_2}}\;  p_2^\mu p_1^\nu + i\,
\frac{c}{m_{V_1}m_{V_2}}\; \epsilon^{\mu\nu\alpha\beta} p_{1\alpha}
p_{2\beta}\right].
\end{eqnarray}
In this basis, the amplitudes with various helicities can be given as
\begin{eqnarray*}
H_{0}=-ax-b(x^2-1), ~~~~~~~H_{\pm}=a\pm \sqrt{x^2-1}~c.
\end{eqnarray*}
where $x=(m^{2}_{B}-m^{2}_{V_1}-m^{2}_{V_2})/(2m_{V_1}m_{V_2})$. In
addition, we can  define the polarization amplitudes as
\begin{eqnarray}
&&A_{0}=H_{0}, ~~~~~~ A_{\parallel}=\frac{1}{\sqrt{2}}(H_{+} +
H_{-})=\sqrt{2}\,a, \non \\
&&A_{\perp}=\frac{1}{\sqrt{2}}(H_{+} -
H_{-})=\sqrt{2}\sqrt{x^2-1}\,c. \label{pol-amp}
\end{eqnarray}
Accordingly, the decay rate expressed by helicity amplitudes for
the VV
mode can be written as
 \begin{eqnarray*}
  \Gamma=\frac{G_F p}{16\pi m^2_{B}}\left( |A_0|^2+ |A_{\parallel}|^2
  +|A_{\perp}|^2\right),
 \end{eqnarray*}
and the polarization fractions can be defined as
\begin{eqnarray}
R_{i}=\frac{|A_{i}|^{2}}{|A_{0}|^2+|A_{\parallel}|^2+|A_{\perp}|^2}.
\label{pf}
\end{eqnarray}
where $i=0$ and $\parallel$ ($\perp$), representing
 the longitudinal and
transverse parallel (perpendicular) components,
respectively, with the relation of $R_{0}+R_{\parallel}+R_{\perp}=1$.
Under CP parities, $R_{0,\parallel}$ are CP-even while $R_{\perp}$
is CP-odd.

 From the hadronic matrix element in Eq. (\ref{hvv}), the
amplitudes $a$, $b$ and $c$ in the framework of the GFA are expressed by
\beq \label{eq:abc}%
a &=& -\tilde{C}_{eff}(m_B+m_{V_{1}})m_{V_{2}}f_{V_{2}}
      A_1^{BV_{1}}(m_{V_{2}}^2), \non \\
b &=& \tilde{C}_{eff}\frac{2m_{V_{1}}m_{V_{2}}^2}
      {m_B+m_{V_1}}f_{V_2}A_2^{BV_1}(m_{V_2}^2), \non \\
c &=& -\tilde{C}_{eff}\frac{2m_{V_1}m_{V_2}^2}{m_B+m_{V_1}}f_{V_2}
      V^{BV_1}(m_{V_2}^2).
\eeq %
where $\tilde{C}_{eff}$ represents the involved WCs and CKM matrix
elements. With the form factors in Eq. (\ref{HQFF}) and the
heavy quark limit, we get that the ratios $r_{b}=b/a$ and $r_{c}=c/a$
 are related. Explicitly, we have
\beq \label{eq:power}%
r_{b}=r_{c}= \frac{m_{D^*}}{m_B}\frac{1}{(1+w)}\approx 0.16,
\eeq %
which are small. From Eqs. (\ref{pol-amp}) and (\ref{pf}),
we find that
the polarization fractions  behave as
\beq %
R_0\sim R_{\parallel},\ \ \ R_{\perp}\sim {\cal O}
\left(\frac{m_{D^*}^2}{m^2_B}\right),
\eeq %
which indicate that the power law in the heavy-heavy decays is
different from the light-light ones, which are expected to be
$R_0\sim 1$, $R_{\parallel}\sim R_{\perp}\sim {\cal
O}(m^2_{V}/m^2_{B})$. Moreover, $R_{\perp}$ is directly related to
$c$ and can be written as
\beq \label{eq:R1}%
R_{\perp}&=&\frac{1}{\Gamma_{0}}(x^2-1)|r_{c}|^{2},
\eeq %
with $\Gamma_{0}=1+(x^2-1)|r_{c}|^2+ \left|
x+(x^2-1)r_{b}\right|^2/2 $. By comparing
to the result in the HQS, we find
an interesting relation
\beq \label{eq:R2}%
\frac{R_{\perp}}{R_{\perp}^{HQS}}\approx \left [ \zeta_{D^*}
  \frac{V(q^2)}{A_1(q^2)} \right ]^2.
\eeq %
where $R^{HQS}_{\perp}= 0.055$ denotes the transverse perpendicular
fraction under the HQS. As a good approximation, the form factor
$A_2$-dependent of $R_{\perp}$ is decoupled. By the relationship,
we can clearly understand the influence of the HQS breaking effects.

\section{Numerical analysis and discussions}

\subsection{Estimations on the annihilation
contributions}\label{sec:anni}

\begin{figure}[htbp]
\includegraphics*[width=3.0in]{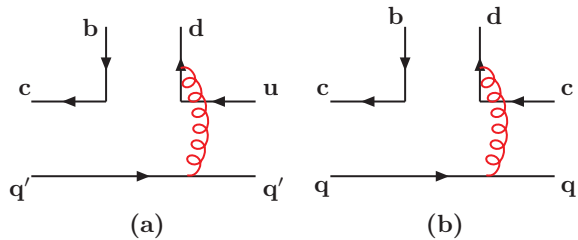}  \caption{
Flavor diagrams for (a) $\bar{B}_d\to D^{+}_{s} K^{-}(J/\Psi
\bar{D}^{0})$ decays while $q^{\prime}=s(c)$ and  (b) $B\to
D^{0}\bar{D}^0 (D^{+}_{s} D^{-}_{s})$ decay while $q=u(s)$. The
gluon attached denotes the nonfactorized effects. }
 \label{fig:anni}
\end{figure}

Since the  annihilation contributions relate to time-like form
factors and there are no direct experimental measurements, we shall
neglect them in our calculations. However, to make sure that the
neglected parts are small, we can connect the processes of $B\to
D^{0} \bar{D}^{0}$ and $B\to D^{+}_{s} D^{-}_{s}$ to the decays
$B^0\to D_s^-K^+$ and $B^0\to J/\psi \bar D^0$,  which are directly
associated with annihilation topologies,
with the experimental data of
$Br(B^0\to D_s^-K^+)=(3.8\pm 1.3)\times 10^{-5}$ \cite{PDG04} and
$Br(B^0\to J/\psi \bar D^0)<1.3\times
10^{-5}$ \cite{JD}, respectively.
By the flavor-diagram analysis, shown in
Fig.~\ref{fig:anni},
except there appears a CKM suppressed factor $V_{cd}\approx
\lambda$ (see Fig.\ref{fig:anni}a) for $D^{0}\bar{D}^{0}$ and
$D^{+}_{s} D^{-}_{s}$ modes, the four modes have the
same decay topologies. Hence,
by assuming that
they have similar hadronic effects, the BRs of $B^0\to D^{0}\bar
D^{0} (D^{+}_s D^{-}_{s})$ could be estimated to be less than ${\cal
O}(10^{-6})$.

To give a detailed analysis, we can include the character of each
mode, governed by the meson distribution amplitudes. For
simplicity, we will concentrate on the leading twist effects and
take the meson
 wave functions to be $\Phi_{D}\propto
f_{D}x(1-x)(1+0.8 (1-2x))$ \cite{PQCD}, $\Phi_{D_s}\propto
f_{D_s}x(1-x)(1+0.3 (1-2x))$ \cite{LU}, $J/\Psi\propto
f_{J/\Psi}x(1-x)(x(1-x)/(1-2.8x(1-x)))^{0.7}$ \cite{BC} and
$\Phi_{K}\propto f_{K} x(1-x)(1+0.51(1-2x)+0.3[5(1-2x)^{2}-1])$
\cite{BBKT},  where
 $x$ is the momentum fraction of the parton inside
the meson and $f_{D,D_s, J/\Psi, K}$ are the decay constants of $D$,
$D_s$, $J/\Psi$ and $K$ mesons, respectively. From these wave
functions, we know that the maximum contributions are from
$x_0\approx (0.35, 0.43,0.5, 0.5)$ for $(D, D_s, J/\Psi, K)$.
 With the information, we can estimate the decay
amplitudes in order of magnitude for $\bar{B}_d\to D^{+}_s K^-
(J/\Psi D^{0})$ and $\bar{B}_{d}\to D^{0} \bar{D}^0 ( D^{+}_s
D^{-}_s)$ as shown in Figs.~\ref{fig:anni}a and
 \ref{fig:anni}b with
$q^{\prime}=s(c)$ and $q=u(s)$, respectively.
Note that
 there exists a chiral suppression in the factorized parts in
annihilation contributions. However,
 we just consider the nonfactorized
effects in the estimations. Therefore, by Fig.~\ref{fig:anni}
with the gluon exchange, the decay amplitude is related to the
propagators of the gluon and the light quark, described by
$1/(k_2+k_3)^2/(k_2+k_3)^2$, where $k_{2(3)}$ denote the momenta
carried by the spectators. For simplicity, we have neglected the
momentum carried by the light quark of the $B$ meson. By the momentum
fraction, the decay amplitude
could satisfy that $A\propto 1/(x_2
x_3)^2 $. Hence, the relative size of the decay amplitudes could be
given approximately as
\begin{eqnarray*}
A(D^{+}_s K^-):A(J/\Psi D^{0}):A(D^{0} \bar{D}^0 ): A(D^{+}_{s}
 \bar{D}^{-}_{s} )\sim \frac{f_{D_s} f_{K}}{(x_{2} x_{3})^2} :
 \frac{f_{J/\Psi} f_{D}}{(x_{2} x_{3})^2}:
 \frac{ \lambda f^{2}_{D} }{(x_{2} x_{3})^2}:
 \frac{ \lambda f^{2}_{D_s}}{(x_{2} x_{3})^2}\; .
\end{eqnarray*}
With the information of maximum contributions, characterized
by $x_0$ for each mode, we get
\begin{eqnarray}
&& A(D^{+}_s K^-):A(J/\Psi D^{0}):A(D^{0} \bar{D}^0 ): A(D^{+}_{s}
\bar{D}^{-}_{s} )\sim
\nonumber\\
&& 1 : \frac{f_{D}f_{J/\Psi}}{f_{D_s}f_K}\left(\frac{0.43}{0.65}
\right)^2:   \frac{\lambda f^2_{D}}{f_{D_s}f_{K}}
\left(\frac{0.5\cdot 0.43 }{0.35^2}\right)^2 : \frac{\lambda
f_{D_s}}{f_{K}} \left(\frac{0.5 }{0.43}\right)^2.
\end{eqnarray}
With the kinetic effects, the ratios of BRs are roughly to be
$BR(D^{+}_{s} K^-): BR(J/\Psi D^{0}): BR(D^0 \bar{D}^{0}):
BR(D^+_{s} \bar{D}^{-}_{s})\sim 1: 0.25 : 0.39: 0.16$. That is, the
BRs of $\bar{B}\to D^{0} \bar{D}^0(D^{+}_{s} D^{-}_{s})$ could be as
large as ${\cal O}(10^{-6})$, which implies that annihilation effects could be
neglected in the discussions on the BR
of the production
for color-allowed two charmful mesons. We note that our estimations are
just based on SD effects and
 at the level of order of magnitude. Clearly,
 direct experimental measurements are needed to confirm our
results.

\subsection{ $\alpha_s$, power corrections and the parametrization of
Isgur-Wise function}

In the HQS limit, the form factors could be related to a single
Isgur-Wise function $\xi(\omega)$ by
$\xi_+=\xi_V=\xi_{A_1}=\xi_{A_3}=\xi(w)$ and $\xi_-=\xi_{A_2}=0$. We
now include the perturbative QCD corrections induced by hard gluon
vertex corrections of $b\to c$ transitions and power corrections in
orders of $1/m_{Q}$ with $Q=b$ and $c$. Consequently,
the form factors can be written as %
\beq %
\xi_i(w)=\left\{ \alpha_i+\beta_i(w)+\gamma_i(w) \right\}~\xi(w).
\eeq %
where $\xi(w)$ is the Isgur-Wise function,
$\alpha_+=\alpha_V=\alpha_{A_1}=\alpha_{A_3}=1$,
$\alpha_-=\alpha_{A_2}=0$ and  $\beta_i(\omega)$ and
$\gamma_{i}(\omega)$ stand for effects of $\alpha_s$ and power
corrections, respectively. Explicitly, for the two-body decays in
our study, $\omega\sim 1.3$ and the values of the other parameters
are summarized as follows \cite{NeubertQCD, NR}:
\beq \label{eq:beta}%
\begin{array}{lll}
  \beta_+=-0.043,     ~~~~~& \beta_-=0.069,       ~~~~~& \beta_V=0.072,  \\
  \beta_{A_1}=-0.067, ~~~~~& \beta_{A_2}=-0.114,  ~~~~~& \beta_{A_3}=-0.028, \\
  \gamma_+=0.015,     ~~~~~& \gamma_-=-0.122,     ~~~~~& \gamma_V=0.224, \\
  \gamma_{A_1}=0.027, ~~~~~& \gamma_{A_2}=-0.093, ~~~~~& \gamma_{A_3}=0.014.
\end{array}
\eeq %
Clearly, the range of their effects is from few percent to 20\%
level. In particular, the power corrections to the form factor $\xi_V$
(or say $V(q^2)$) are the largest, about 20\%. The resultant is also
consistent with other QCD approaches, such as the
constitute quark model (CQM) \cite{MS} and the
light-front (LF) QCD \cite{CCH}.

After taking care of the corrections, the remaining unknown is the
Isgur-Wise function. To determine it, we adopt a linear
parametrization to be $\xi(w)=1-\rho^2_{H}(w-1)$ for the transition
$B\to H$, where $\rho^2_{H}$ is called the slope parameter. We use
the BRs of semileptonic $B\to D^{(*)}\ell^{-} \bar{\nu}_{\ell}$
decays to determine $\rho^2_{H}$. We note that the values of
$\rho^2_{D}$ and $\rho^{2}_{D^*}$ are not the same as those in $D$
and $D^*$ decays. In our approach, the difference is from higher
orders and power corrections. Hence, with $V_{cb}=0.0416$ and
$BR(B\to D^{(*)} \ell^{-} \bar{\nu}_{\ell})= 2.15\pm 0.22 ( 6.5\pm
0.5)\%$ \cite{PDG04}, we obtain $\rho^2_{D}=0.90\pm 0.06$ and
$\rho^2_{D^*}=1.09\pm 0.05$. Since the errors of $\rho^2_{H}$ are
small, we will only use the central values in our numerical results.

\subsection{ Results for BRs and polarization fractions}

To get the numerical estimations, the input values for the relevant
parameters are taken to be as follows \cite{PDG04,NS, CKM}: 
\beq &&f_D=0.20,~~\ f_{D^*}=0.23,~~ f_{D_s}=0.24,~~
f_{D_s^*}=0.275~\GeV;
\nonumber\\
&&V_{cd}=-\lambda,~~~~~~~~~~ V_{cs}=1-\lambda^2/2,~~~~~~~~~
V_{cb}=A\lambda^2, \non \\
&&V_{td}=\lambda |V_{cb}|R_t e^{-i\phi_{1}}, ~~~~~
V_{ts}=-A\lambda^2,~~~~~
V_{tb}=1, \non \\
&&A=0.83,~~~~~\lambda=0.224,~~~~~\phi_1=23.4^{\rm o},~~~~~R_t=0.91;
\nonumber\\
&& m_u=0.005,\ m_d=0.01,\ m_s=0.15,\ m_c =1.5,\ m_b=4.5~\GeV.
\eeq %
 Note that the numerical results are insensitive to
light quark masses. As to the WCs, we adopt the formulas up to
one-loop corrections presented in Ref. \cite{GFA2} and set $\mu=2.5$
GeV. As mentioned early, since the nonfactorized contributions are
grouped into $N^{\rm eff}_{c}$, the color number in Eq.
(\ref{effwc}) will be regarded as a variable. To display their
effects, we take the values of $N^{\rm eff}_{c}=2,\; 3,\; 5$ and
$\infty$.

By following the factorized formulas shown in Appendixes, we present
the BRs with various $N^{\eff}_{c}$ in Tables \ref{tab:PP},
\ref{tab:PV} and \ref{tab:VV} for $PP$, $PV(VP)$ and $VV$ modes,
respectively. In order to accord with the experimental data, our
predictions of the BRs are given as the CP-averaged values. For
comparisons, we also calculate the results in terms of the form
factors given by the CQM and LF, which are displayed in
Table~\ref{tab:QCD_models}. Since  the CPAs are quite similar in
different models,  in Table~\ref{tab:QCD_models} we just show the
results in our approach. As to the polarization fractions, we
present them in Table~\ref{tab:fraction}. Therein, to understand the
influence of the HQS breaking effects, we separate the results to be
HQS and HQS$_{\rm I(II)}$, representing the HQS results and those
with $\alpha_{s}$ ($\alpha_{s}$+power) corrections, respectively.

\begin{table}[hptb]
\caption{\label{tab:PP} BRs (in unit of $10^{-3}$) for $B\to PP$
decays with $\rho^2_{D}=0.90$.}
\begin{ruledtabular}
\begin{tabular}{cccccc}
mode & $N^{\rm eff}_c=2$ & $N^{\rm eff}_c=3$  & $N^{\rm eff}_c=5$ &
$N^{\rm eff}_{c}=\infty$  & Exp. \\ \hline %
$\bar{B}^0\to D^+ D_s^-$  & 7.26  & 8.25  & 9.06  & 10.46
 & $8\pm 3$  \cite{PDG04} \\ \hline %
$B^-\to D^0 D_s^- $       & 7.85  & 8.94  & 9.82  & 11.34
 & $13\pm 4$ \cite{PDG04} \\ \hline %
$\bar{B}^0\to D^+D^-$     & 0.28  & 0.31  & 0.34  &  0.40
 & $0.321\pm 0.057\pm 0.048$  \cite{BELLE_DD} \\ \hline %
$B^-\to  D^0 D^-$         & 0.30  & 0.33  & 0.37  &  0.43
 & $0.562\pm 0.082\pm 0.065$ \cite{BELLE_DD} \\ %
\end{tabular}
\end{ruledtabular}
\end{table}
%
\begin{table}[hptb]
\caption{\label{tab:PV} BRs (in unit of $10^{-3}$) for $B\to PV
(VP)$ decays with $\rho^2_{D^{(*)}}=0.90(1.09)$.}
\begin{ruledtabular}
\begin{tabular}{cccccc}
mode & $N^{\rm eff}_c=2$ & $N^{\rm eff}_c=3$  & $N^{\rm eff}_c=5$ &
$N^{\rm eff}_{c}=\infty$  & Exp.  \\ \hline %
$\bar{B}^0\to D^+ D_s^{*-}$ &  9.52 & 10.80 & 11.84 & 13.62
 & $10\pm 5$ \cite{PDG04} \\ \hline %
$\bar{B}^0\to D^{*+} D_s^-$ &  6.78 &  7.67 &  8.41 &  9.66
 & $10.7\pm 2.9$ \cite{PDG04} \\ \hline %
$B^-\to  D^0 D_s^{*-}$      & 10.35 & 11.73 & 12.87 & 14.79
 & $9\pm 4$ \cite{PDG04} \\ \hline %
$B^-\to  D^{*0} D_s^-$      &  7.37 &  8.34 &  9.14 & 10.49
 & $12\pm 5$ \cite{PDG04} \\ \hline %
$\bar B^0\to D^{*+}D^-$     &  0.25 &  0.29 &  0.32 &  0.36 \\ \hline %
$\bar B^0\to D^+D^{*-}$     &  0.37 &  0.42 &  0.46 &  0.53 \\ \hline %
$\bar B^0\to D^{*+}D^-$     &  0.62 &  0.71 &  0.78 &  0.89
 & $0.93\pm 0.15$ \cite{PDG04} \\  %
$~~~~~+D^+D^{*-}$          &       &       &       &
 &    \\ \hline
$B^-\to  D^0 D^{*-}$        &  0.40 &  0.45 &  0.50 &  0.57
 & $0.459\pm 0.072\pm 0.056$ \cite{BELLE_DD} \\ \hline %
$B^-\to D^{*0} D^{-}$       &  0.28 &  0.31 &  0.34 &  0.39 \\ %
\end{tabular}
\end{ruledtabular}
\end{table}
%
\begin{table}[hptb]
\caption{\label{tab:VV} BRs (in unit of $10^{-3}$) for $B\to VV$
decays with $\rho^2_{D^*}=1.09$.}
\begin{ruledtabular}
\begin{tabular}{cccccc}
mode & $N^{\rm eff}_c=2$ & $N^{\rm eff}_c=3$  & $N^{\rm eff}_c=5$ &
$N^{\rm eff}_{c}=\infty$  & Exp.  \\ \hline %
$\bar B^0\to D^{*+} D_s^{*-}$ & 22.52 & 25.51 & 27.98 & 32.19
 & $19\pm 5$ \cite{PDG04};   \\
                              &       &       &       &
 & $18.8\pm 0.9\pm 1.7$ \cite{BABAR_DD} \\ \hline %
$B^-\to D^{*0} D_s^{*-}$      & 24.44 & 27.69 & 30.37 & 34.93
 & $27\pm 10$ \cite{PDG04} \\ \hline %
$\bar B^0\to D^{*+}D^{*-}$    &  0.87 &  0.91 &  0.99 &  1.14
 & $0.81\pm 0.08\pm 0.11$ \cite{BELLE_DD}; \\
                              &       &       &       &
 & $0.87\pm 0.18$ \cite{PDG04}  \\ \hline %
$B^-\to D^{*0} D^{*-}$        &  0.81 &  0.98 &  1.08 &  1.24 \\ %
\end{tabular}
\end{ruledtabular}
\end{table}

\begin{table}[hptb]
\caption{\label{tab:QCD_models} BRs (in unit of $10^{-3}$)
with $N^{\rm
eff}_{c}=3$ and the form factors calculated in the CQM, LF and HQSC,
while the CPA is only shown in our approach (HQSC).}
\begin{ruledtabular}
\begin{tabular}{ccccc}
Mode                        & CQM   & LF    & HQSC  & $A_{CP}(\%)$ \\ \hline %
$\bar{B}^0\to D^+ D_s^-$    &  9.70 & 10.33 &  8.25 & $-0.2$ \\ \hline %
$B^-\to D^0 D_s^- $         & 10.58 & 11.26 &  8.94 & $-0.2$ \\ \hline %
$\bar{B}^0\to D^+D^-$       &  0.38 &  0.40 &  0.31 & $ 2.5$ \\ \hline %
$B^-\to  D^0 D^-$           &  0.42 &  0.44 &  0.33 & $ 2.5$ \\ \hline \hline %
$\bar{B}^0\to D^+ D_s^{*-}$ & 12.49 & 11.42 & 10.80 & $-0.1$ \\ \hline %
$\bar{B}^0\to D^{*+} D_s^-$ &  9.19 &  8.50 &  7.67 & $ 0.0$ \\ \hline %
$B^-\to  D^0 D_s^{*-}$      & 13.65 & 12.47 & 11.73 & $-0.1$ \\ \hline %
$B^-\to  D^{*0} D_s^-$      & 10.02 &  9.27 &  8.34 & $ 0.0$ \\ \hline %
$\bar B^0\to D^{*+}D^-$     &  0.36 &  0.33 &  0.29 & $ 0.2$ \\ \hline %
$\bar B^0\to D^+D^{*-}$     &  0.37 &  0.45 &  0.42 & $ 0.9$ \\ \hline %
$\bar B^0\to D^{*+}D^-$     &  0.73 &  0.78 &  0.71 & $ 0.6$ \\ %
 $~~~~~+D^+D^{*-}$          &       &       &       &        \\ \hline
$B^-\to D^0 D^{*-}$         &  0.54 &  0.49 &  0.45 & $ 0.9$ \\ \hline %
$B^-\to D^{*0} D^-$         &  0.39 &  0.36 &  0.31 & $ 0.2$ \\ \hline \hline %
$\bar B^0\to D^{*+}D_s^{*-}$& 28.78 & 27.09 & 25.51 & $-0.1$ \\ \hline %
$B^-\to D^{*0} D_s^{*-}$    & 31.37 & 29.52 & 27.69 & $-0.1$ \\ \hline %
$\bar B^0\to D^{*+}D^{*-}$  &  1.06 &  0.99 &  0.91 & $ 0.9$ \\ \hline %
$B^-\to D^{*0} D^{*-}$      &  1.16 &  1.08 &  0.98 & $ 0.9$ \\ %
\end{tabular}
\end{ruledtabular}
\end{table}

\begin{table}[hptb]
\caption{\label{tab:fraction} Polarization fractions in various QCD
approaches, where HQS$_{\rm I}$ and HQS$_{\rm II}$
 denote the results with
$\alpha_{s}$
and $\alpha_{s}$ + power corrections, respectively. }
\begin{ruledtabular}
\begin{tabular}{cccccc}
Mode  & CQM  &  LF  & HQS  & HQS$_{\rm I}$  & HQS$_{\rm II}$    \\  \hline %
$\bar B^0\to D^{*-} D_s^{*+}$
 & $R_0=0.523$  &  $R_0=0.512$ & $R_0=0.515$
 & $R_0=0.517$  & $R_0=0.512$   \\  %
 & $R_{\perp}=0.069$ & $R_{\perp}=0.077$ & $R_{\perp}=0.055$
 & $R_{\perp}=0.070$ & $R_{\perp}=0.093$  \\ \hline %
$B^-\to D^{*0} D_s^{*-}$
 & $R_0=0.524$  &  $R_0=0.512$ & $R_0=0.515$
 & $R_0=0.517$ & $R_0=0.512$   \\  %
 & $R_{\perp}=0.070$ & $R_{\perp}=0.078$ & $R_{\perp}=0.055$
 & $R_{\perp}=0.070$ & $R_{\perp}=0.093$  \\ \hline %
$\bar B^0\to D^{*+}D^{*-}$
 & $R_0=0.547$  &  $R_0=0.535$  & $R_0=0.538$
 & $R_0=0.540$ & $R_0=0.535$   \\  %
 & $R_{\perp}=0.069$ & $R_{\perp}=0.077$ & $R_{\perp}=0.055$
 & $R_{\perp}=0.070$ & $R_{\perp}=0.092$  \\ \hline %
$B^-\to  D^{*0} D^{*-}$
 & $R_0=0.547$  &  $R_0=0.535$  & $R_0=0.538$
 & $R_0=0.541$ & $R_0=0.535$   \\  %
 & $R_{\perp}=0.069$ & $R_{\perp}=0.077$& $R_{\perp}=0.055$
 & $R_{\perp}=0.070$ & $R_{\perp}=0.092$  \\  %
\end{tabular}
\end{ruledtabular}
\end{table}

We now present our discussions on the results as follows:

(1) The non-factorizable contributions are not dominant for
color-allowed two charmed-meson decays. According to the
classification in Refs. \cite{GFA1, GFA2}, the decay modes displaced
in Tables \ref{tab:PP}, \ref{tab:PV} and \ref{tab:VV} belong to
class I, which are dominated by the external $W$-emission. The
leading decay amplitudes are proportional to the effective
coefficient $a_1$, which is stable against the variation of
$N_c^{\eff}$. Thus, the predicted branching ratios are  insensitive
to  $N_c^{\eff}$. This means that annihilation contributions and
FSIs, neglected in the GFA, are sub-leading contributions. On the
other hand, by varying $N_c^{\eff}$ from $3$ to $2$, or $3$ to
$\infty$, the branching ratios change by about 10-20\%, which should
be the same order as annihilation and FSI effects. From Tables
\ref{tab:PP}, \ref{tab:PV} and \ref{tab:VV}, there are no obvious
deviations of the theoretical predictions from the experimental data
within the present errors. It is also interesting to note that
$N_c^{\eff}=\infty$ is not excluded by experiments if considering the
uncertainties of decay constants and from factors. Thus, the large
$N_c$ limit as a mechanism of factorization is not disfavored yet.

(2) The main uncertainties of theory come from the decay constants
and form factors. Because the decay amplitudes are proportional to
decay constants, it is clear that the theoretical predictions can be
changed with different values of the decay constants. For instance,
the branching ratio is ${\rm BR(our~ result)} \times
\left(\frac{f_{D_s}}{0.24} \right)^2$ for $\bar B^0\to D^+D_s^-$.
The recent experiment ${\rm BR}(\bar B^0\to D^{*+} D_s^{*-})=18.8\pm
0.9\pm 1.7$ seems to favor a lower $f_{D_s^*}\approx 0.24$ than our
choice of $0.275$. However, this point has to be checked by other
processes. For the form factors, the predictions of BRs in our
approach are slightly lower than those in other two approaches (CQM
and LF). The present experiment data can not distinguish which model is
more preferred. More precise data are necessary. Another place to
test different approaches is through the transverse polarization
$R_{\perp}$. From Table~\ref{tab:fraction}, $R_{\perp}$ is predicted
to be $0.07, ~0.08$ and $0.09$ in the CQM, LF and HQET,
respectively. The larger prediction in the HQET is due to $\alpha_s$
corrections. Except the model-dependent calculation of power
corrections in different approaches, one advantage of the HQET is
that it permits the calculations of perturbative QCD corrections
systematically.

(3) The penguin effects can not be neglected in $B\to PP$ decays.
By using the decay amplitudes in Appendixes, the definitions of
hadronic effects in Eqs. (\ref{xpp}) and (\ref{xpv}) and the condition of
$\varepsilon_{i}(p_i)\cdot p_{i}=0$, we know that the effects of
penguin ($P$) to tree ($T$) level, denoted by $P/T$, for $PP$, $VP$
and $VV$ modes are proportional to $(a^{\rm eff(c)}_{4}+2a^{\rm
eff(c)}_{6} {\cal R})/a^{\rm eff}_{1}$, $(a^{\rm eff(c)}_{4}-2a^{\rm
eff(c)}_{6} {\cal R'})/a^{\rm eff}_{1}$ and $a^{\rm
eff(c)}_{4}/a^{\rm eff}_{1}$, respectively, where ${\cal
R}=m^{2}_{D}/[(m_{c}+m_{d})(m_{b}-m_{c})]$ and  ${\cal
R'}=m^{2}_{D}/[(m_{c}+m_{d})(m_{b}+m_{c})]$ and the CKM matrix
elements have been canceled due to $|V_{tb}V^*_{ts}|\approx
|V_{cb}V^{*}_{cs}|$ and $|V_{tb}V^{*}_{td}|\approx |V_{cb}
V^{*}_{cd}|$. The situations in the $PV$ modes are
 the same as those in the $VV$
modes due to the vector meson being factorized out from the $B\to P$
transition. Since the WCs $a^{\rm eff(c)}_{4}$ and $a^{\rm
eff(c)}_{6}$ have the same sign, we see clearly that penguin effects
in the $PP$ modes are larger than those in the $VV$ modes; however,
due to the cancelation between $a^{\rm eff(c)}_{4}$ and $a^{\rm
eff(c)}_6$, penguin effects could be neglected in $B\to VP$ decays.
Hence, the ratios $|P/T|$ for $PP$, $VP$ and $VV(PV))$ are around
$15\%$, $0\%$ and $4\%$, respectively. For the $PP, VV(PV)$ modes,
our predictions are consistent with the results in Refs. \cite{CY,
KKLM}. Note that an 4\% penguin contribution was obtained for the
$VP$ modes in \cite{CY}. The difference is due to that they used a
lower charm quark mass ($m_c=0.95~\GeV$) than ours. For all the
decay modes, the electroweak penguin contributions can be negligible
(less than 1\%).

(4) Without FSIs, we find that the BRs in the neutral and charged modes
have the following relationships:
\begin{eqnarray*}
\frac{1}{\tau_{B^0}}BR(D^{(*)+} D^{(*)-}_{s}) &\approx&
\frac{1}{\tau_{B^+}}BR( D^{(*)0} D^{(*)-}_{s}), \\
\frac{1}{\tau_{B^0}}BR(D^{(*)+} D^{(*)-})&\approx &
\frac{1}{\tau_{B^+}}BR( D^{(*)0} D^{(*)-}).
\end{eqnarray*}
In addition, the decays with nonstrangeness charmed mesons are
Cabibbo-suppressed compared to the decays with the $D^{(*)}_{s}$
emission and they satisfy
\begin{eqnarray}
BR(B\to D^{(*)} D^{(*)})\approx \frac{ f^{2}_{D^{(*)}} }{
f^{2}_{D^{(*)}_{s} }} \lambda^2  BR(B\to D^{(*)} D^{(*)}_{s}).
\label{equal}
\end{eqnarray}
Clearly, if large deviations from the equalities in Eq.
(\ref{equal}) are observed in experiments,
they should arise from FSIs. Of course, if the BRs of
$\bar B^0 \to D^{(*)0} \bar{D}^{(*)0}$ and $\bar B^0 \to
D^{(*)+}_{s} D^{(*)-}_{s} $ with ${\cal O}(10^{-4})$  are seen,
it will be another hint for FSIs \cite{EFP}.

(5) For the decay amplitude, we write
\begin{eqnarray}
A=T + Pe^{i\theta_{W}} e^{i\delta},
\end{eqnarray}
where $T$ and $P$ represent tree and penguin amplitudes, and
 we have chosen the convention such that $T$ and $P$ are real
numbers and $\theta_W$ and $\delta$ are the CP weak and strong
phases, respectively.  From Eq. (\ref{cpa}), the CPA
can be described by
\begin{eqnarray}
A_{CP}={ 2(P/T)\sin\delta \sin\theta_{W} \over 1+ (P/T)^2+2(P/T)
\cos\delta \cos\theta_{W}}.
\end{eqnarray}
According to the discussions in (1), the maximum CPAs
in $PP$, $PV$ and $VV(VP)$ are expected to be around $26\%$,
$0\%$ and $8\%$, respectively. However, in  $B\to D^{(*)}
D^{(*)}$ decays, due to $|\theta_{W}|=|\phi_{1}|$, if we take
$\delta=90^{o}$ and $\phi=23.4^{o}$, the maximum CPAs for $PP$ and
$VV(VP)$ modes are $10.3\%$ and $1.6\%$, respectively.
Clearly, in the SM, the CPA
with ${\cal O}(10\%)$ can be reached in $B\to DD$ decays. Due to
the associated CKM matrix element being $V_{ts}\approx -A\lambda^2$,
there are no CPAs in $B\to D^{(*)}D^{(*)}_s$ decays. In the GFA,
since the strong phases mainly arise from the one-loop corrections
which are usually small, our results on CPAs, shown in
Table~\ref{tab:QCD_models}, are all at a few percent level.
Therefore, if the CPAs of ${\cal O}(10\%)$ are found in $\bar B^0\to
D^{+} D^{-}$ and $\bar B^+\to D^{0} D^{-}$ decays, we can conclude
the large effects of the strong phase are from FSIs.

(6) As discussed before, we know that in two charmful decays the
polarization fractions satisfy $R_{\perp}<<R_{0}\sim R_{\parallel}$.
The current experimental data are:
$R_0=0.52\pm 0.05$ \cite{PDG04} for $B^0\to D^{*+} D_s^{*-}$ and
$R_0=0.57\pm 0.08\pm 0.02$, $R_{\perp}=0.19\pm 0.08\pm 0.01$
\cite{BELLE_DD} and $R_{\perp}=0.063\pm 0.055\pm 0.009$
\cite{BABAR_DD} for $B^0\to D^{*+}D^{*-}$.
We can see that the experimental measurements
support the power-law relation. To estimate how large $R_{\perp}$
can be in theory, we use the relationship in Eq. (\ref{eq:R2})
and the form factors in Eq. (\ref{ffv}) and  we obtain
\begin{eqnarray}
\frac{R_{\perp}}{R_{\perp}^{HQS}}\approx \left [
 \frac{1+\beta_V+\gamma_V}{1+\beta_{A_1}+\gamma_{A_1}} \right ]^2.
\end{eqnarray}
With the values in Eq. (\ref{eq:beta}) and $R^{HQS}_{\perp}=0.055$,
we get $R_{\perp} \approx 10\%$. The detailed numerical values
can be found in Table~{\ref{tab:fraction}}. Interestingly, for the
$\bar B^0\to D^{*+}D^{*-}$ decay, the estimated result is close to
the upper limit of $R_{\perp}=0.063\pm 0.055\pm 0.009$ observed by
BABAR \cite{BABAR_DD} but close to the lower limit of
$R_{\perp}=0.19\pm 0.08\pm 0.01$ observed by BELLE \cite{BELLE_DD}.
We note that our results are different from the PQCD predictions
in which $R_{\perp}\sim 0.06$ \cite{LM}. From our results,
we can conjecture that if large $R_{\perp}$, say around $20\%$, is
observed,  large contributions should arise from FSIs.

\section{Conclusions}

We have presented a detailed study of B decaying into two
charmed-mesons in the generalized factorization approach. The
penguin contributions have also been taken into account. If the
final states are both pseudoscalar mesons, the ratio of penguin and
tree contributions is about 10\% in the decay amplitude. The direct
CP violating asymmetries have been estimated to be a few percent.
For the $B^0\to D^{*+}D^-, ~D^+D^{*-}, ~ D^{*+}D^{*-}$ decays, the
``penguin pollution" is weaker than that in the $D^+D^-$ mode. Thus,
these modes provide cleaner places to cross-check the value of
 $sin 2\beta$ measured in the $B^0\to J/\psi K$ decays. The weak
annihilation contributions have been found to be small. We have
proposed to test the annihilation effects in annihilation-dominated
processes of $B^0\to D^{(*)0}\bar D^{(*)0}$ and $D^{(*)+}_s
D^{(*)-}_{s}$.

We have performed a comprehensive test of the factorization in the
heavy-heavy B decays. The predictions of branching ratios in theory
are consistent with the experimental data within the present level.
The variations of branching ratios with the effective color number
$N_c^{\eff}$ show that the soft FSIs are not dominant. However, we
cannot make the conclusion that they are negligible. Their effects
can be of order 10-20\% for branching ratios as indicated from the
variation of $N_c^{\eff}$. Since the soft divergences are not
canceled in the non-factorizable corrections, this may indicate that
the strong interactions at low energy either become weak or are
suppressed by some unknown parameters (such as $N_c$ in the large
$N_c$ theory). If the factorization is still a working concept in
the heavy-heavy decays, there must be some non-perturbative
mechanisms which prefer the factorization of a large-size
charmed-meson from an environment of ``soft cloud''. A relevant
comment on the necessity of non-perturbative QCD justification can
be found in \cite{LLW}.

The polarization structure in the heavy-heavy decays has shown that
the transverse perpendicular polarization fraction $R_{\perp}$ is
the smallest while the other two are comparable in size. This
structure follows from the QCD dynamics in the heavy quark limit. We
have found one relation between the transverse perpendicular
polarization fraction and the ratios of form factors, in particular
$V(q^2)/A_1(q^2)$. The corrections to the heavy quark limit give an
enhancement of $R_{\perp}$ from 0.055 to about 0.09. Since the FSIs
are not significant, we do not expect that FSIs can change our prediction
of $R_{\perp}$ substantially. If future measurements confirm
$R_{\perp}\sim 0.2$ as the recent measurement by BELLE, it will be
difficult to explain within the HQET and the factorization
hypothesis.

In conclusion, our study has shown that the factorization works well in
B meson heavy-heavy decays at present. More precise experimental
data are desired to give a better justification. For theory, to
explain the mechanism of factorization in the heavy-heavy decays is of
high interest. The measurement of the transverse perpendicular
polarization provides important information on the size of the heavy
quark symmetry breaking or the possibility of large non-factorizable
effects.

\vspace{0.5cm} {\bf Acknowledgments}\\
We thank Hai-Yang Cheng and Yu-Kuo Hsiao for many valuable
discussions. This work is supported in part by the National Science
Council of R.O.C. under Grant \#s: NSC-93-2112-M-006-010 and
NSC-93-2112-M-007-014.

\appendix

\section*{$B\to PP$ decays}
\begin{eqnarray}
A(\bar{B}^0\to D^{+} D^{-}_{s} )&=& V_{cb}V^{*}_{cs} a^{\rm eff}_{1}
X^{(BD,D_s)}_{1}-V_{tb}V^{*}_{ts} \left[ \left(a^{\rm eff(c)}_{4}
X^{(BD,D_s)}_{1}-2a^{\rm eff(c)}_6 X^{(BD,D_s)}_2\right) \right.\nonumber \\
&&+ \left.\left( a^{\rm eff(d)}_{4} Y^{(B,DD_s)}_1-2a^{\rm
eff(d)}_{6} Y^{(B,DD_s)}_3\right) \right],
\end{eqnarray}

\begin{eqnarray}
A(\bar{B}^0\to D^{+}_{s} D^{-}_{s} )&=&  V_{cb}V^{*}_{cd}   a^{\rm
eff}_{2} Y^{(B,D_sD_s)}_{1} -V_{tb}V^{*}_{td} \left[ (a^{\rm
eff(s)}_{3}+ a^{\rm
eff(c)}_{3}) Y^{(B,D_sD_s)}_1 \right. \nonumber \\
&&+ \left. (a^{\rm eff(s)}_{5}+ a^{\rm eff(c)}_{5}) Y^{(B,D_sD_s)}_2
\right],
\end{eqnarray}

\begin{eqnarray}
A(B^-\to D^{0} D^{-}_{s} )&=& V_{cb}V^{*}_{cs} a^{\rm eff}_{1}
X^{(BD,D_s)}_{1}+V_{ub}V^{*}_{us}a^{\rm eff}_{1} Y^{(B,DD_s)}_{1}
-V_{tb}V^{*}_{ts} \left[ \left(a^{\rm eff(c)}_{4} X^{(BD,D_s)}_{1}
\right. \right. \nonumber \\
&&- \left. \left. 2a^{\rm eff(c)}_6 X^{(BD,D_s)}_2\right) + \left(
a^{\rm eff(u)}_{4} Y^{(B,DD_s)}_1-2a^{\rm eff(u)}_{6}
Y^{(B,DD_s)}_3\right) \right],
\end{eqnarray}

\begin{eqnarray}
A(\bar{B}^0\to D^{+} D^{-} )&=& V_{cb}V^{*}_{cd} \left[a^{\rm
eff}_{1} X^{(BD,D)}_1+ a^{\rm eff}_{2}
Y^{(B,DD)}_{1}\right]-V_{tb}V^{*}_{td} \left[ \left( a^{\rm
eff(c)}_{4} X^{(BD,D)}_{1} \right. \right. \nonumber \\
&&-\left. 2a^{\rm eff(c)}_{6} X^{(BD,D)}_{2}\right)+(a^{\rm
eff(d)}_{4}+a^{\rm eff(d)}_{3}+ a^{\rm
eff(c)}_{3}) Y^{(B,DD)}_1 \nonumber \\
&&+ \left. (a^{\rm eff(d)}_{5}+ a^{\rm eff(c)}_{5}) Y^{(B,DD)}_2
-2a^{\rm eff(d)}_{6} Y^{(B,DD)}_3\right],
\end{eqnarray}

\begin{eqnarray}
A(\bar{B}^0 \to D^{0} \bar{D}^{0} )&=& \left( V_{cb}V^{*}_{cd}  +
V_{ub}V^{*}_{ud} \right) a^{\rm eff}_{2} Y^{(B,DD)}_{1}
-V_{tb}V^{*}_{td} \left[ (a^{\rm eff(u)}_{3}+ a^{\rm
eff(c)}_{3}) Y^{(B,DD)}_1 \right. \nonumber \\
&&+ \left. (a^{\rm eff(u)}_{5}+ a^{\rm eff(c)}_{5}) Y^{(B,DD)}_2
\right],
\end{eqnarray}

\begin{eqnarray}
A(B^- \to D^{0} D^{-} )&=& V_{cb}V^{*}_{cd} a^{\rm eff}_{1}
X^{(BD,D)}_{1}+V_{ub}V^{*}_{ud} a^{\rm eff}_{1} Y^{(B,DD)}_1
-V_{tb}V^{*}_{td} \left[ \left( a^{\rm
eff(c)}_{4} X^{(BD,D)}_{1} \right. \right. \nonumber \\
&&- \left. \left. 2a^{\rm eff(c)}_{6} X^{(BD,D)}_{2}\right)+a^{\rm
eff(d)}_{4}Y^{(B,DD)}_1 -2a^{\rm eff(d)}_{6} Y^{(B,DD)}_3\right].
\end{eqnarray}

\section{$B\to PV(VP)$ decays}

\begin{eqnarray}
A(\bar{B}^0 \to D^{+} D^{*-}_{s} )&=& V_{cb}V^{*}_{cs} a^{\rm
eff}_{1} X^{(BD,D^*_s)}_1-V_{tb}V^{*}_{ts} \left[ a^{\rm eff(c)}_{4}
X^{(BD,D^*_s)}_{1} \right.\nonumber \\
&&+ \left.\left( a^{\rm eff(d)}_{4} Y^{(B,DD^*_s)}_1-2a^{\rm
eff(d)}_{6} Y^{(B,DD^*_s)}_3\right) \right],
\end{eqnarray}

\begin{eqnarray}
A(\bar{B}^0 \to D^{*+} D^{-}_{s} )&=& V_{cb}V^{*}_{cs} a^{\rm
eff}_{1} X^{(BD^*,D_s)}_1-V_{tb}V^{*}_{ts} \left[ \left(a^{\rm
eff(c)}_{4}
X^{(BD^*,D_s)}_{1}-2a^{\rm eff(c)}_6 X^{(BD^*,D_s)}_2\right) \right.\nonumber \\
&&+ \left.\left( a^{\rm eff(d)}_{4} Y^{(B,D^*D_s)}_1-2a^{\rm
eff(d)}_{6} Y^{(B,D^*D_s)}_3\right) \right],
\end{eqnarray}

\begin{eqnarray}
A(\bar{B}^0 \to D^{+}_{s} D^{*-}_{s} )&=&  V_{cb}V^{*}_{cd} a^{\rm
eff}_{2} Y^{(B,D_sD^*_s)}_{1} -V_{tb}V^{*}_{td} \left[ (a^{\rm
eff(s)}_{3}+ a^{\rm
eff(c)}_{3}) Y^{(B,D_sD^*_s)}_1 \right. \nonumber \\
&&+ \left. (a^{\rm eff(s)}_{5}+ a^{\rm eff(c)}_{5})
Y^{(B,D_sD^*_s)}_2 \right],
\end{eqnarray}

\begin{eqnarray}
A(\bar{B}^0 \to D^{*+}_{s} D^{-}_{s} )&=&  V_{cb}V^{*}_{cd} a^{\rm
eff}_{2} Y^{(B,D^*_sD_s)}_{1} -V_{tb}V^{*}_{td} \left[ (a^{\rm
eff(s)}_{3}+ a^{\rm
eff(c)}_{3}) Y^{(B,D^*_sD_s)}_1 \right. \nonumber \\
&&= \left. (a^{\rm eff(s)}_{5}+ a^{\rm eff(c)}_{5})
Y^{(B,D^*_sD_s)}_2 \right],
\end{eqnarray}

\begin{eqnarray}
A(B^- \to D^{0} D^{*-}_{s} )&=& V_{cb}V^{*}_{cs} a^{\rm eff}_{1}
X^{(BD,D^*_s)}_1+V_{ub}V^{*}_{us}a^{\rm eff}_{1} X^{(B,DD^*_s)}_{1}
-V_{tb}V^{*}_{ts} \left[ \left(a^{\rm eff(c)}_{4} X^{(BD,D^*_s)}_{1}
\right. \right. \nonumber \\
&&+ \left.  \left( a^{\rm eff(u)}_{4} Y^{(B,DD^*_s)}_1-2a^{\rm
eff(u)}_{6} Y^{(B,DD^*_s)}_3\right) \right],
\end{eqnarray}

\begin{eqnarray}
A(B^- \to D^{*0} D^{-}_{s} )&=& V_{cb}V^{*}_{cs} a^{\rm eff}_{1}
X^{(BD^*,D_s)}_1+V_{ub}V^{*}_{us}a^{\rm eff}_{1} Y^{(B,D^*D_s)}_{1}
-V_{tb}V^{*}_{ts} \left[ \left(a^{\rm eff(c)}_{4} X^{(BD^*,D_s)}_{1}
\right. \right. \nonumber \\
&&- \left. \left. 2a^{\rm eff(c)}_6 X^{(BD^*,D_s)}_2\right) + \left(
a^{\rm eff(u)}_{4} Y^{(B,D^*D_s)}_1-2a^{\rm eff(u)}_{6}
Y^{(B,D^*D_s)}_3\right) \right],
\end{eqnarray}

\begin{eqnarray}
A(\bar{B}^0 \to D^{+} D^{*-} )&=& V_{cb}V^{*}_{cd} \left[a^{\rm
eff}_{1} X^{(BD,D^*)}_1+ a^{\rm eff}_{2}
Y^{(B,DD^*)}_{1}\right]-V_{tb}V^{*}_{td} \left[ a^{\rm
eff(c)}_{4} X^{(BD,D^*)}_{1} \right.  \nonumber \\
&&+ (a^{\rm eff(d)}_{4}+a^{\rm eff(d)}_{3}+ a^{\rm eff(c)}_{3})
Y^{(B,DD^*)}_1 + (a^{\rm eff(d)}_{5}+ a^{\rm eff(c)}_{5})
Y^{(B,DD^*)}_2  \nonumber \\
&&- \left. 2a^{\rm eff(d)}_{6} Y^{(B,DD^*)}_3\right],
\end{eqnarray}

\begin{eqnarray}
A(\bar{B}^0 \to D^{*+} D^{-} )&=& V_{cb}V^{*}_{cd} \left[a^{\rm
eff}_{1} X^{(BD^*,D)}_1+ a^{\rm eff}_{2}
Y^{(B,D^*D)}_{1}\right]-V_{tb}V^{*}_{td} \left[ \left( a^{\rm
eff(c)}_{4} X^{(BD^*,D)}_{1} \right. \right. \nonumber \\
&&- \left. 2a^{\rm eff(c)}_{6} X^{(BD^*,D)}_{2}\right)+(a^{\rm
eff(d)}_{4}+a^{\rm eff(d)}_{3}+ a^{\rm
eff(c)}_{3}) Y^{(B,D^*D)}_1 \nonumber \\
&&+ \left. (a^{\rm eff(d)}_{5}+ a^{\rm eff(c)}_{5}) Y^{(B,D^*D)}_2
-2a^{\rm eff(d)}_{6} Y^{(B,D^*D)}_3\right],
\end{eqnarray}

\begin{eqnarray}
A(\bar{B}^0 \to D^{0} \bar{D}^{*0} )&=& \left( V_{cb}V^{*}_{cd}  +
V_{ub}V^{*}_{ud} \right) a^{\rm eff}_{2} Y^{(B,DD^*)}_{1}
-V_{tb}V^{*}_{td} \left[ (a^{\rm eff(u)}_{3}+ a^{\rm
eff(c)}_{3}) Y^{(B,DD^*)}_1 \right. \nonumber \\
&&+ \left. (a^{\rm eff(u)}_{5}+ a^{\rm eff(c)}_{5}) Y^{(B,DD^*)}_2
\right],
\end{eqnarray}

\begin{eqnarray}
A(\bar{B}^0 \to D^{*0} \bar{D}^{0} )&=& \left( V_{cb}V^{*}_{cd}  +
V_{ub}V^{*}_{ud} \right) a^{\rm eff}_{2} Y^{(B,D^*D)}_{1}
-V_{tb}V^{*}_{td} \left[ (a^{\rm eff(u)}_{3}+ a^{\rm
eff(c)}_{3}) Y^{(B,D^*D)}_1 \right. \nonumber \\
&&+ \left. (a^{\rm eff(u)}_{5}+ a^{\rm eff(c)}_{5}) Y^{(B,D^*D)}_2
\right],
\end{eqnarray}

\begin{eqnarray}
A(B^- \to D^{0} D^{*-} )&=& V_{cb}V^{*}_{cd} a^{\rm eff}_{1}
X^{(BD,D^{*})}_1+V_{ub}V^{*}_{ud} a^{\rm eff}_{1} Y^{(B,DD^*)}_1
-V_{tb}V^{*}_{td} \left[  a^{\rm
eff(c)}_{4} X^{(BD,D^*)}_{1} \right. \nonumber \\
&&+ \left. a^{\rm eff(d)}_{4}Y^{(B,DD^*)}_1 -2a^{\rm eff(d)}_{6}
Y^{(B,DD^*)}_3\right],
\end{eqnarray}

\begin{eqnarray}
A(B^- \to D^{*0} D^{-} )&=& V_{cb}V^{*}_{cd} a^{\rm eff}_{1}
X^{(BD^{*},D)}_1+V_{ub}V^{*}_{ud} a^{\rm eff}_{1} Y^{(B,D^{*}D)}_1
-V_{tb}V^{*}_{td} \left[ \left( a^{\rm
eff(c)}_{4} X^{(BD^{*},D)}_{1} \right. \right. \nonumber \\
&&- \left. \left. 2a^{\rm eff(c)}_{6}
X^{(BD^{*},D)}_{2}\right)+a^{\rm eff(d)}_{4}Y^{(B,D^{*}D)}_1
-2a^{\rm eff(d)}_{6} Y^{(B,D^*D)}_3\right].
\end{eqnarray}

\section{$B\to VV$ decays}

\begin{eqnarray}
A(\bar{B}^0 \to D^{*+} D^{*-}_{s} )&=& V_{cb}V^{*}_{cs} a^{\rm
eff}_{1} X^{(BD^*,D^*_s)}-V_{tb}V^{*}_{ts} \left[ a^{\rm eff(c)}_{4}
X^{(BD^*,D^*_s)}_{1} \right.\nonumber \\
&&+ \left.\left( a^{\rm eff(d)}_{4} Y^{(B,D^*D^*_s)}_1-2a^{\rm
eff(d)}_{6} Y^{(B,D^*D^*_s)}_3\right) \right],
\end{eqnarray}

\begin{eqnarray}
A(\bar{B}^0 \to D^{*+}_{s} D^{*-}_{s} )&=&  V_{cb}V^{*}_{cd}
a^{\rm eff}_{2} Y^{(B,D^*_sD^*_s)}_{1} -V_{tb}V^{*}_{td} \left[
(a^{\rm eff(s)}_{3}+ a^{\rm
eff(c)}_{3}) Y^{(B,D^*_sD^*_s)}_1 \right. \nonumber \\
&&+ \left. (a^{\rm eff(s)}_{5}+ a^{\rm eff(c)}_{5})
Y^{(B,D^*_sD^*_s)}_2 \right],
\end{eqnarray}

\begin{eqnarray}
A(B^- \to D^{*0} D^{*-}_{s} )&=& V_{cb}V^{*}_{cs} a^{\rm eff}_{1}
X^{(BD^*,D^*_s)}+V_{ub}V^{*}_{us}a^{\rm eff}_{1}
Y^{(B,D^*D^*_s)}_{1} -V_{tb}V^{*}_{ts} \left[a^{\rm eff(c)}_{4}
X^{(BD^*,D^*_s)}_{1}
\right. \nonumber \\
&&+ \left.   \left( a^{\rm eff(u)}_{4} Y^{(B,D^*D^*_s)}_1-2a^{\rm
eff(u)}_{6} Y^{(B,D^*D^*_s)}_3\right) \right],
\end{eqnarray}

\begin{eqnarray}
A(\bar{B}^0 \to D^{*+} D^{*-} )&=& V_{cb}V^{*}_{cd} \left[a^{\rm
eff} X^{(BD^*,D^*)}+ a^{\rm eff}_{2}
Y^{(B,D^*D^*)}_{1}\right]-V_{tb}V^{*}_{td} \left[ \left( a^{\rm
eff(c)}_{4} X^{(BD^*,D^*)} \right. \right. \nonumber \\
&&+(a^{\rm eff(d)}_{4}+a^{\rm eff(d)}_{3}+ a^{\rm eff(c)}_{3})
Y^{(B,D^*D^*)}_1 + (a^{\rm eff(d)}_{5}+ a^{\rm eff(c)}_{5})
Y^{(B,D^*D^*)}_2 \nonumber \\
&& \left. -2a^{\rm eff(d)}_{6} Y^{(B,D^*D^*)}_3\right],
\end{eqnarray}

\begin{eqnarray}
A(\bar{B}^0 \to D^{*0} \bar{D}^{*0} )&=& \left( V_{cb}V^{*}_{cd}  +
V_{ub}V^{*}_{ud} \right) a^{\rm eff}_{2} Y^{(B,D^*D^*)}_{1}
-V_{tb}V^{*}_{td} \left[ (a^{\rm eff(u)}_{3}+ a^{\rm
eff(c)}_{3}) Y^{(B,D^*D^*)}_1 \right. \nonumber \\
&&+ \left. (a^{\rm eff(u)}_{5}+ a^{\rm eff(c)}_{5}) Y^{(B,D^*D^*)}_2
\right],
\end{eqnarray}

\begin{eqnarray}
A(B^- \to D^{*0} D^{*-} )&=& V_{cb}V^{*}_{cd} a^{\rm eff}
X^{(BD^*,D^*)}+V_{ub}V^{*}_{ud} a^{\rm eff}_{1} Y^{(B,D^*D^*)}_1
-V_{tb}V^{*}_{td} \left[ a^{\rm
eff(c)}_{4} X^{(BD^*,D^*)} \right.  \nonumber \\
&&\left. +a^{\rm eff(d)}_{4}Y^{(B,DD)}_1 -2a^{\rm eff(d)}_{6}
Y^{(B,DD)}_3\right].
\end{eqnarray}

\end{document}